\documentclass[12pt]{article}
\usepackage{ae,amssymb,amsbsy,amsthm,amsfonts,turnstile}
\usepackage{tensor}
\usepackage{cite}
\usepackage[english]{babel} 
\usepackage[hmarginratio=1:1,top=32mm,columnsep=20pt]{geometry} 
\newcommand{\cir}[1]{\left(#1\right)}		
\newcommand{\cuad}[1]{\left[#1\right]}		
\newcommand{\corch}[1]{\left\{#1\right\}}	

\newcommand{\dif}[2]{\frac{\delta#1}{\delta#2}}		
\newcommand{\vc}[1]{\textbf{#1}}		
\newcommand{\punto}[2]{(\vc{#1}\cdot\vc{#2})}	
\newcommand{\cruz}[2]{(\vc{#1}\times\vc{#2})}	
\newcommand{\bsym}[1]{\boldsymbol{#1}}		
\newcommand{\dn}[2]{\frac{d#1}{d#2}}		
\title{\textbf {Generalized Papapetrou's equations of motion for an extended test body within static and isotropic metrics}}
\author{William Almonacid and Leonardo Casta\~neda\\\\
\small Observatorio Astron\'omico Nacional, Universidad Nacional de Colombia,\\
\small Edificio 413, Bogot\'a, Colombia\\
\footnotesize E-mail: \texttt{waalmonacidg@unal.edu.co, lcastanedac@unal.edu.co}
}
\date{}
\begin{document}
\maketitle

\begin{abstract}
Applying Dixon's general equations of motion for extended bodies, we compute the Papapetrou's equations for an extended test body on static and isotropic metrics. We incorporate the force and the torque terms which involve multipole moments, beyond dipole moment, from the energy-momentum tensor. We obtain the vector form equations for both Corinaldesi-Papapetrou and Tulczyjew-Dixon spin supplementary conditions. An expanded effective mass, including interactions between the structure of the body and the gravitational field, is also found.
\end{abstract}
\pagebreak
\section{Introduction}

It is well known that the motion of a test particle, without any structure, is performed through the geodesics of the background field. Nevertheless, even if the body doesn't disturb the spacetime, when the structure of the body is involved, the equation of motion deviates from that of the geodesic. Corrections to the motion of test bodies, due to its multipolar structure, were first derived by Mathisson \cite{mathisson1,mathisson2} and Papapetrou \cite{papapetrou}. Into their works, covariant general-relativistic equations of motion were obtained and then were written in a vector form within the Schwarzschild field \cite{corinaldesi,barker}. However, the analyses focused on spinning-bodies, getting equations up to the dipole approximation. Additional studies under different gravitational fields have been carried out, but up to the same approximation \cite{wald,carmeli,semerak,hartl,plyatsko,hackmann}.\\

Although more general approaches to extended bodies have been already develo\-ped \cite{dixon1,dixon2,dixon3,harte,steinhoff1}, a complete equations of motion for test bodies, beyond the dipole level, can shed light on the contribution of the structure to the motion of the body in external field, and also on the coupling between spacetime and mass-energy distribution. Progress has been made in this area with the introduction of quadrupole dynamics, by setting particular quadrupole models \cite{bini,steinhoff2}.\\

The principal goal of this paper is to compute Papapetrou-like form equations for an extended test body in static and isotropic metrics beyond the pole-dipole approximation, starting from the Dixon's general equations. Firstly, we briefly summarize the basic elements of Dixon's approach to extended bodies. From the Dixon's equations we get the covariant equations in a Papapetrou form in section 3. In section 4 we present the set of equations in vector form, adopting isotropic coordinates and applying both Corinaldesi-Papapetrou and Tulczyjew-Dixon spin supplementary conditions. A discussion about the results are finally presented in section 5.
\section{Dixon's equations of motion}

The Papapetrou's equations can be regarded as a pole-dipole approximation of an extended body equations of motion \cite{dixon1}. Within the framework of General Relativity an extended body is outlined by a space-time four-dimensional region called world tube and its structure is governed by the momentum-energy distribution. Inside the world tube a world line $\gamma$ is chosen to represent the motion of the body \cite{dixon2}. If an extended test body is considered, the self-forces are neglected, i.e., the influence of the body on the gravitational field is despised; therefore the structure of the spacetime is dominated by an external field.\\

Dixon \cite{dixon3} wrote the equations of motion in terms of the 4-momentum $p^\nu$ and spin tensor $S^{\mu\nu}$, which are defined by integrals of
the particle's stress-energy tensor $T^{\mu\nu}$ over an arbitrary spacelike hypersurface $\Sigma$ (the set of spacelike hypersurfaces represent a foliation ${\Sigma_s}$ of the manifold \cite{kriele}):

\begin{equation}\label{momentum1}
 p^\kappa(z,\Sigma) \equiv \int_\Sigma T^{\alpha\beta}K_\alpha^{\ \kappa} d\Sigma_\beta,
\end{equation}

\begin{equation}\label{angular1}
 S^{\kappa\lambda}(z,\Sigma) \equiv 2 \int_\Sigma T^{\alpha\beta} H_\alpha^{[\kappa}\Omega^{\lambda]} d\Sigma_\beta,
\end{equation}

where $z(s)$ parametrize the worldline $\gamma$ of the body, $\Omega$ is the Synge's world function \cite{synge}, $K^\kappa_{\ \alpha}$ and $H^\kappa_{\ \alpha}$ are bitensors known as Jacobi propagators \cite{dixon1}. The equations of motion for an extended body are then

\begin{equation}\label{fuerza1}
 \dif{p^\nu}{s} - \frac{1}{2} S^{\kappa\lambda}v^\mu R\indices{_\kappa_\lambda_\mu^\nu} = F^\nu,
\end{equation}

\begin{equation}\label{torque1}
 \dif{S^{\kappa\lambda}}{s} - 2 p^{[\kappa}v^{\lambda]} = L^{\kappa\lambda},
\end{equation}

where $\dif{}{s}=v^\mu\nabla_\mu$ and $v^\mu$ is the 4-velocity, i.e., the tangent to the body's worldline. The terms denoted by $F^\nu$ and $L^{\kappa\lambda}$ are known as the force and torque and both represent the coupling between the internal structure of the body and the external field, by means of a multipole expansion which has been computed by Dixon \cite{dixon3} and Harte \cite{harte} by different ways (see appendix).\\

There is an arbitrariness in the choice of $z$, which is linked to the definition of center of mass in General Relativity. The uniqueness of this definition is shown in the Pryce \cite{pryce}, Beiglb\"ock \cite{beiglbock} and Schattner \cite{schattner} works. The arbitrariness can be fixed by selecting $z_0$ inside the body such that

\begin{equation}\label{CM8}
 u_\alpha(z)S^{\alpha\beta}(z) = 0,
\end{equation}

for some unit timelike (future-pointing) vector $u^\alpha$. An unique $u^\alpha$ can be picked out at each point by the condition

\begin{equation}\label{CM7}
 p^{[\alpha}u^{\beta]} = 0, \qquad u_\beta u^\beta = -1,
\end{equation}

and the world line may be parametrized in such a way that

\begin{equation}\label{parametro}
 u_\kappa v^\kappa=-1.
\end{equation}

This corresponds to choosing $ds$, at each point of $\gamma$, to be the elapsed time interval in the frame in which the 3-momentum is zero \cite{dixon4}.\\

From (\ref{CM7})

\begin{equation}\label{CM9}
 p^\kappa = M u^\kappa,
\end{equation}

where $M$ is a positive quantity interpreted as the total mass of the body, which is not constant in general. It is important to keep in mind that $u^\kappa$ differs from the body's velocity $v^\kappa$. The former is the so called dynamical velocity and the latter the cinema\-tical velocity. A special attribute of $u^\alpha$ is to allow the fixing of the foliation ${\Sigma_s}$ of the spacetime, since the hypersurfaces are composed of all geodesics through $z$ orthogonal to $u^\alpha$. Therefore, (\ref{CM9}) sets up ${\Sigma_s}$ as the chosen one by zero momentum observers.\\

The arbitrariness discussed above is related to an essential feature of the Dixon equations, which is the freedom of a specific definition of the representing world line $\gamma$. This freedom manifests itself in the fact that the system (\ref{fuerza1}-\ref{torque1}) is not closed and the number of unknowns exceeds the number of equations. Therefore the world line can be determined arbitrarily from physical considerations. The consequence is the spin supplementary condition (\ref{CM8}), which picks out a unique world line that is identified as the center of mass \cite{costa,karpov}. In this paper we choose two: the Corinaldesi-Papapetrou condition



\begin{equation}\label{CSScp}
S^{i0}=0,
\end{equation}

which chooses the center of mass of the body measured by the rest frame of the central attractting body; and the Tulczyjew-Dixon condition \cite{dixons,tulczyjew}

\begin{equation}\label{CSStd}
 p_\nu S^{\mu\nu} = 0,
\end{equation}

where the center of mass is the point around which its dipole moment of mass va\-nishes at zero 3-momentum observers.\\

\section{Generalized Papapetrou's equations of motion}

In this section we will get the Papapetrou's equations of motion from the general equations introduced by Dixon. We incorporate the force and torque terms. The equations are presented in covariant form and they are independent of a particular choice of center of mass.\\

Consider the general equations of motion (\ref{fuerza1},\ref{torque1}), contracting (\ref{torque1}) with $u_\nu$ and by reducing (\ref{CM7}),(\ref{parametro}) and (\ref{CM9}), we obtain:

\begin{equation}\label{3spin1}
 u_\nu \dif{S^{\mu\nu}}{s} = -p^\mu + Mv^\mu + u_\nu L^{\mu\nu}.
\end{equation}

Hence the moment may be rewritten as

\begin{equation}\label{3moment1}
 p^\mu = M v^\mu - u_\sigma \dif{S^{\mu\sigma}}{s} + u_\sigma L^{\mu\sigma},
\end{equation}

which is replaced in (\ref{torque1}) to get

\begin{equation}\label{3spin2}
 \dif{S^{\mu\nu}}{s} + 2u_\sigma v^{[\nu}\dif{S^{\mu]\sigma}}{s} + 2u_\sigma v^{[\mu}L^{\nu]\sigma} - L^{\mu\nu}=0.
\end{equation}

This expression is consistent with the covariant form of the Papapetrou's spin equation. They are strictly identical when the torque is neglected. In this case we would be dealing with a pole-dipole approximation.\\

The antisymmetry of the spin tensor reduces (\ref{3spin2}) to six independent equations. However, choosing a spin supplementary condition allows us to relate the additional components of $S^{\mu\nu}$ to each other, thus the amount of lineal independent equations decrease. Then it is useful to write the spatial components of (\ref{3spin2}) as

\begin{equation}\label{3spin3}
 \dif{S^{ij}}{s} + \frac{v^j}{v^0}\dif{S^{0i}}{s} - \frac{v^i}{v^0}\dif{S^{0j}}{s} - \frac{2}{v^0}v^{[i}L^{j]0} - L^{ij} = 0.
\end{equation}

On the other hand, equation (\ref{3moment1}) displays a substantial difference between the li\-neal momentum and the velocity of the body, which arises from its structure ($S^{\mu\nu}$ and $L^{\mu\nu}$).\\

The components of (\ref{3moment1}) read

\begin{eqnarray}\label{3moment2}
 p^0 = Mv^0 - u_\sigma \dif{S^{0\sigma}}{s} + u_\sigma L^{0\sigma},\\
 p^i = Mv^i - u_0\dif{S^{i0}}{s} - u_j\dif{S^{ij}}{s} + u_\sigma L^{i\sigma}.
\end{eqnarray}

Substituting (\ref{3spin3}) into (\ref{3moment1}), the lineal momentum reduces to

\begin{equation}\label{3moment3}
p^\mu = M_*v^\mu - \frac{1}{v^0}\dif{S^{0\mu}}{s} + \frac{1}{v^0}L^{0\mu},
\end{equation}

where $M_*$ represents an effective mass associated with the mass of the body and an energetic component induced by the interaction between the multipole structure of the body and the spacetime curvature. This mass is defined by

\begin{equation}\label{3masa1}
 M_* = M + M_s + M_L,
\end{equation}

with

\begin{equation}\label{3masa1a}
M_s=\frac{u_\sigma}{v^0}\dif{S^{\sigma0}}{s}\qquad {\rm and} \qquad M_L= \frac{u_\sigma}{v^0}L^{0\sigma}.
\end{equation}

From (\ref{3moment3}) we obtain the following equation of motion:

\begin{equation}\label{3moment4}
 \dif{M_*v^\mu}{s} - \dif{p^\mu}{s} - \dif{}{s}\cir{\frac{1}{v^0}\dif{S^{0\mu}}{s}} + \dif{}{s}\cir{\frac{1}{v^0}L^{0\mu}} = 0.
\end{equation}

The differentiation of the momentum can be obtained from (\ref{fuerza1}) by employing the symmetry relations of the angular moment and the Riemann tensor, thus

\begin{equation}\label{3momentum}
 \dif{p^\mu}{s} = -S^{\kappa\lambda}v^\nu\cir{\partial_\kappa\Gamma^\mu_{\nu\lambda} + \Gamma^\mu_{\sigma\kappa}\Gamma^\sigma_{\nu\lambda}} + F^\mu.
\end{equation}

When higher multipole moments ($F^\mu$ and $L^{\mu\nu}$) are neglected, the equation (\ref{3moment3}) reproduces the Papapetrou's equation for the lineal momentum, with $M_*=M + M_s$. Where $M_s$ represents an energy associated with spin-orbit coupling.

\section{Equations of motion in isotropic coordinates}

The adoption of isotropic coordinates to write the equations of motion of the extended test body allows us to set the equations in vector form, which leads  to understanding the role of some terms like the effective mass, the spin-orbit coupling and the multipole moments.\\

In these coordinates the metric is

\begin{equation}\label{EImetric}
 g_{00} = -A(r),\qquad y\qquad g_{ij} = \delta_{ij} - \frac{(1-B(r))}{r^2}x^i x^j,
\end{equation}

where the spatial coordinates are abbreviated by the vector $\vc{r}$ with $r^2=x^ix^j\delta_{ij}$. The non-zero Christoffel symbols read

\begin{eqnarray}\label{3Chffel}
  \Gamma^0_{0i} &=& \frac{\mu'_A}{2r}x^i,\qquad \Gamma^i_{00} = \frac{A'}{2rB}x^i\nonumber\\\\
  \Gamma^i_{jk} &=& \corch{\frac{1}{2r^2}\cuad{\mu'_B+\frac{2(1-B)}{rB}}x^jx^k - \frac{(1-B)}{rB}\delta_{jk}}\frac{x^i}{r},\nonumber
\end{eqnarray}

where the prime symbol means differenciation with respect to $r$, $\mu'_A=A'/A$ and $\mu'_B=B'/B$.\\ 

\subsection{Equations of motion \textsl{under} Corinaldesi-Papapetrou supplementary condition}

As we exposed above the spin supplementary condition closes the system (\ref{3spin2}) and (\ref{3momentum}). If the Corinaldesi-Papapetrou condition (\ref{CSScp}) is selected, the equation (\ref{CSScp}) yields

\begin{equation}\label{3dspin1}
 \dif{S^{ij}}{s}=\dn{S^{ij}}{s} + \Gamma^i_{\kappa\lambda}v^\kappa S^{\lambda j} + \Gamma^j_{\kappa\lambda}v^\kappa S^{i\lambda},
\end{equation}

\begin{equation}\label{3dspin2}
 \dif{S^{0i}}{s}=\Gamma^0_{\kappa\lambda}v^\kappa S^{\lambda i}.
\end{equation}

Substituting (\ref{3dspin1}) and (\ref{3dspin2}) into equation (\ref{3spin3}) we obtain

\begin{equation}\label{CPspin}
 \dif{S^{ij}}{s} + \Gamma^0_{\kappa\lambda}\frac{v^\kappa}{v^0}\cir{v^jS^{\lambda i}-v^iS^{\lambda j}} - \frac{2}{v^0}L^{0[i}v^{j]} - L^{ij} = 0,
\end{equation}

which is the equation for the angular momentum in the Schwarzschild rest frame. The equation of motion would be given by (\ref{3moment4}) and (\ref{3momentum}), such that

\begin{eqnarray}\label{CPmoment}
  \dif{(M_*v^\mu)}{s} - \dif{}{s}\cir{\frac{1}{v^0}\Gamma^0_{\kappa\lambda}v^\kappa S^{\lambda\mu}} &+& S^{\kappa\lambda}v^\nu\cir{\partial_\kappa\Gamma^\mu_{\nu\lambda} + \Gamma^\mu_{\sigma\kappa}\Gamma^\sigma_{\nu\lambda}}\nonumber\\
                  &+& F^\mu + \dif{}{s}\cir{\frac{1}{v^0}L^{0\mu}} = 0.
\end{eqnarray}

In addition, the spin supplementary condition reduces the independent components of the spin tensor to three, such that a spin vector can be defined by

\begin{equation}\label{spinvec}
S^{k} = \frac{1}{2}\epsilon_{ijm}\delta^{km} S^{ij},
\end{equation}

with $\epsilon_{ijk}$ the Levi-Civita symbol. Hence, the equation of motion for the angular momentum (\ref{CPspin}), with (\ref{EImetric}) and (\ref{3Chffel}), reduces to

\begin{eqnarray}\label{CPspin1}
 \dot{\vc{S}} + \frac{1}{2r}(\mu'_B-\mu'_A)(\vc{r}\cdot\vc{v})\vc{S} &+& \frac{(1-B)}{r^2B}(\vc{r}\cdot\vc{S})\vc{v} + \frac{\mu'_A}{2r}(\vc{v}\cdot\vc{S})\vc{r}\\ 
			&-&\frac{1}{2r^3}\cuad{\mu'_B+\frac{2(1-B)}{rB}}(\vc{r}\cdot\vc{v})(\vc{r}\cdot\vc{S})\vc{r} - \bsym{\tau} = 0,\nonumber
\end{eqnarray}

where the dot means differenciation with respect to the parameter $s$ fixed by (\ref{parametro}).\\

The vector $\bsym{\tau}$ represents the torque contribution. It satisfies

\begin{equation}\label{3torque1}
 \bsym{\tau} = \frac{1}{2}\epsilon_{ijm}\cir{\frac{1}{2}L^{ij}+\frac{1}{v^0}L^{0[i}v^{j]}}\delta^{km}.
\end{equation}

This expression shows a coupling between the velocity of the representative center of mass and the higher multipole moments in $L^{\mu\nu}$.\\ 

By applying the spin supplementary condition into equation (\ref{CPmoment}) we obtain the equation of motion,

\begin{equation}\label{CPmoment0}
 \dn{}{s}(M_*v^0)+M_*\lambda^0-F^0 = 0
\end{equation}

and

\begin{eqnarray}\label{CPmoment1}
 \dn{}{s}(M_*\vc{v})+M_*\bsym{\lambda} &+& f(r)\cuad{\vc{S}\cdot(\vc{r}\times\vc{v})}\vc{r} + g(r)(\vc{r}\cdot\vc{v})(\vc{r}\times\vc{S})\nonumber\\
 &+& h(r)(\vc{v}\times\vc{S}) + \vc{F} - \frac{\mu'_A}{2r}(\vc{r}\times\bsym{\tau}) - \bsym{\varsigma}= 0.
\end{eqnarray}

Where

\begin{eqnarray}\label{lambdas1}
  \lambda^0 &=& c\frac{\mu'_A}{r}\punto{r}{v}\dot{t},\\
  \bsym{\lambda} &=& \frac{1}{2r}\corch{\frac{c^2A'}{B}\dot{t}^2-\frac{2(1-B)}{rB}|\vc{v}|^2+\frac{1}{r^2}\cuad{\mu'_B+\frac{2(1-B)}{rB}}\punto{r}{v}^2}\vc{r},
\end{eqnarray}

\begin{equation}\label{3torque2}
 \bsym{\varsigma}=\dif{}{s}\cir{\frac{1}{v^0}L^{0i}},
\end{equation}

and $f$, $g$ y $h$ are functions of the coordinates related to the gravitational potentials. They are defined by \cite{almonacid2}

\begin{eqnarray}\label{CPfunction}
 f(r) &=&-\frac{1}{2r^4B}(2-2B+\mu'_Br),\qquad h(r)=\frac{1}{2r^2B}(2-2B-\mu'_Ar),\\\nonumber\\
 g(r) &=& -\frac{1}{2r^4B}(2-2B-\mu'_Ar)-\frac{1}{4r^3}\cir{2\frac{A''}{A}r+2\mu'_B-\mu'_A\mu'_Br-\mu'^2_Ar}.
\end{eqnarray}

The last three terms in (\ref{CPmoment1}) represent higher multipole moments contribution.\\

When a pole-dipole particle is considered the equations (\ref{CPmoment0}) and (\ref{CPmoment1}) yield conservatives quantities. In that case, for example, if we cancel $F^0$ into (\ref{CPmoment0}) we have

\begin{equation}\label{3consv1}
 \dn{}{s}(AM_*v^0)=0,
\end{equation}

with $M_*=M+M_s$. From the equation (\ref{3consv1}) one may define the energy integral by $E=AM_*v^0$.\\

Finally, we can express (\ref{3masa1}) as

\begin{equation}\label{3masa2}
M_s=\frac{\mu'_A}{2rM}(\vc{r}\times\vc{p})\cdot\vc{S}.
\end{equation}

Therefore (\ref{3masa1}) gets the characteristic form of the spin-orbit coupling. It is important to keep in mind that the momentum vector is not equal to the product of mass and velocity, and its behavior is determined by the equation (\ref{fuerza1}), which has the vector form

\begin{eqnarray}\label{CPmoment2}
 \dif{\vc{p}}{s}-f(r)\cuad{\vc{S}\cdot\cruz{r}{v}}\vc{r}&+&\frac{(2-2B+rB')}{2r^2B}\punto{r}{v}\cruz{r}{S}\nonumber\\
 &-&\frac{(1-B)}{r^2B}\cruz{v}{S}-\vc{F}=0.
\end{eqnarray}

\subsection{Equations of motion \textsl{under} Tulczyjew-Dixon supplementary condition}\label{TD}

By means of the Tulczyjew-Dixon supplementary condition (\ref{CSStd}) the independent components of the spin tensor are reduced to three, such that

\begin{equation}\label{CSStd1}
 S^{0i} = \frac{P_j}{P_0}S^{ij}.
\end{equation}

Hence, the definition (\ref{spinvec}) can be used again. Aplying index exercise with the metric (\ref{EImetric}), the equation (\ref{CSStd1}) reduces to

\begin{equation}\label{3spin5}
 S^{0i} = -\frac{1}{Ap^0}\cuad{(\vc{p}\times\vc{S})-\frac{(1-B)}{r^2}(\vc{r}\cdot\vc{p})(\vc{r}\times\vc{S})}^i.
\end{equation}

The equation of motion for the center of mass is gotten aplying the spin supplementary condition to equations (\ref{fuerza1}-\ref{torque1}), then

\begin{equation}\label{vcm}
 M v^\mu = p^\mu - u_\nu L^{\mu\nu} - \frac{S^{\mu\nu}\cuad{MF_\nu-\frac{1}{2}S^{\kappa\lambda}\cir{p^\tau - u_\sigma L^{\tau\sigma}}R_{\kappa\lambda\nu\tau}}}
									{M^2+\frac{1}{4}S^{\kappa\lambda}S^{\tau\nu}R_{\kappa\lambda\tau\nu}}.
\end{equation}

When equation (\ref{vcm}) is substituted in (\ref{torque1}) the kinetic term only depends on the torque and quadratric terms of the spin, such a way the equation of motion for the angular momentum will give

\begin{eqnarray}\label{3spin6}
 \dot{\vc{S}} &-& \frac{\mu'_A}{2rBM_*} \cuad{\vc{r}\times(\vc{p}\times\vc{S})-\frac{(1-B)}{r^2}(\vc{r}\cdot\vc{p})(\vc{r}\times(\vc{r}\times\vc{S}))}\nonumber\\
 &+& \frac{1}{2r^3}\cuad{\mu'_B+\frac{2(1-B)}{rB}}(\vc{r}\cdot\vc{v})\cuad{r^2\vc{S}-(\vc{r}\cdot\vc{S})\vc{r}}\\
 &+& \frac{(1-B)}{r^2B}\cuad{(\vc{r}\cdot\vc{S})\vc{v}-(\vc{r}\cdot\vc{v})\vc{S}} + \vc{N} = 0,\nonumber
\end{eqnarray}

with

\begin{equation}\label{3masa3}
 M_*=\frac{p^0}{Ã—v^0},
\end{equation}

as a result of (\ref{3moment3}). The contribution of the torque and quadratric terms of the spin are gathered at $\vc{N}$, such that

\begin{equation}\label{3Nspin}
 \vc{N} =  \frac{1}{2}\epsilon_{ijm}\corch{\frac{u_\nu p^{[i}L^{j]\nu}}{M} + \frac{p^{[i}S^{j]\nu}}{M}\frac{\cuad{MF_\nu-\frac{1}{2}S^{\kappa\lambda}\cir{p^\tau - u_\sigma L^{\tau\sigma}}R_{\kappa\lambda\nu\tau}}}
				             {M^2+\frac{1}{4}S^{\kappa\lambda}S^{\tau\nu}R_{\kappa\lambda\tau\nu}} - L^{ij}}\delta^{km}.
\end{equation}

Making use of the identity

\begin{equation}\label{vecident}
 \vc{a}\times(\vc{b}\times\vc{c})= (\vc{a}\cdot\vc{c})\vc{b} - (\vc{a}\cdot\vc{b})\vc{c},
\end{equation}

(\ref{3spin6}) reduces to

\begin{eqnarray}\label{TDspin}
 \dot{\vc{S}} &+& \frac{1}{2r}\cuad{\frac{\mu'_A}{M_*}(\vc{r}\cdot\vc{p}) + \mu'_B(\vc{r}\cdot\vc{v})}\vc{S} + \frac{(1-B)}{r^2B}(\vc{r}\cdot\vc{S})\vc{v}
 - \frac{\mu'_A}{2rBM_*}(\vc{r}\cdot\vc{S})\vc{p}\nonumber\\\\
 &+& \frac{1}{2r^3}\cuad{\frac{\mu'_A}{M_*}\frac{(1-B)}{B}(\vc{r}\cdot\vc{p}) - \cir{\mu'_B + 
  \frac{2(1-B)}{rB}}(\vc{r}\cdot\vc{v})}(\vc{r}\cdot\vc{S})\vc{r} + \vc{N} = 0.\nonumber
\end{eqnarray}

On the other hand, the equation of motion (\ref{3moment4}) includes quadratric terms of the spin and satisfies

\begin{eqnarray}\label{TDmoment}
 \dn{}{s}(M\vc{v}) &+& M\bsym{\lambda} + \frac{1}{M_*}\tilde{f}(r)\cuad{(\vc{r}\times\vc{p})\cdot\vc{S}}\vc{r} + \frac{\mu'_A}{2rBM_*}(\vc{p}\times\vc{S})\nonumber\\
 &-& \frac{mu'_A}{2r^3BM_*}(1-B)(\vc{r}\cdot\vc{p})(\vc{r}\times\vc{S}) + \tilde{g}(r)\cuad{(\vc{r}\times\vc{v})\cdot\vc{S}}\vc{r}\\
 &+& \frac{1}{2r^3}\cuad{\mu'_B+\frac{2(1-B)}{rB}}(\vc{r}\cdot\vc{v})(\vc{r}\times\vc{S}) - \frac{(1-B)}{r^2B}(\vc{v}\times\vc{S}) + \bsym{\Psi} = 0,\nonumber
\end{eqnarray}

with

\begin{equation}\label{3Psi}
\bsym{\Psi} = F^i - \dif{}{s}\corch{u_\nu L^{i\nu} + \frac{S^{i\nu}\cuad{MF_\nu-\frac{1}{2}S^{\kappa\lambda}\cir{p^\tau - u_\sigma L^{\tau\sigma}}R_{\kappa\lambda\nu\tau}}}{M^2+\frac{1}{4}S^{\kappa\lambda}S^{\tau\nu}R_{\kappa\lambda\tau\nu}}}
\end{equation}

and

\begin{eqnarray}\label{TDfuncion}
 \tilde{f}(r) &=& \frac{1}{4r^3B}\cir{2\frac{A''}{A}r-2\mu'_A-\mu'_A\mu'_Br-\mu'^2_Ar}\nonumber\\\\
 \tilde{g}(r) &=& \frac{1}{2r^4}\cir{2-\frac{2}{B}+\frac{\mu'_B}{B}r}.
\end{eqnarray}

It is important to remember that the effective mass $M_*$ is defined by (\ref{3masa1}). However, with the current supplementary condition, $M_s$ is delimeted by the variation of $S^{0i}$ in (\ref{3spin5}). Hence, the effective mass is function of the derivatives of momentum and angular momentum. In addition $\vc{r}$ is not equal to its analogue in the Corinaldesi-Papapetrou supplementary condition, because in each case it represents the vector pointing the center of mass in different reference frames \cite{barker}.

\section{Discussion}

By employing the extended bodies dynamics, we derived the Papapetrou-like form equations of motion for an extended test body in an arbitrary external field, and then we wrote them within static and isotropic metric. The general equations agree with those obtained for a pole-dipole particle by Papapetrou \cite{papapetrou} and the vector form equations written in the chosen metric are consistent with those given for Schwarzschild field in \cite{corinaldesi} and \cite{barker}. The latter are shown in \cite{almonacid}.\\

We extend the effective mass proposed by Papapetrou, by including in (\ref{3masa1}) the new term $M_L$, which represents the energy associated with the interaction between the test body structure and the gravitational field, coupled with the dynamical velocity. In addition, the characteristic form of spin-orbit interaction energy, symbolized by $M_s$, is preserved. We also find additional terms in the equations of motion, related to force and torque, which are disengaged from those of pole-dipole approximation. In (\ref{CPspin1}) the multipolar contribution to the spin evolution involves spatial components of the torque and torque-orbit coupling, which are represented by $\bsym{\tau}$ in (\ref{3torque1}). Whereas the last three terms displayed in (\ref{CPmoment1}) describes quadrupolar and beyond orders contributions.\\

In the Tulczyjew-Dixon condition context, the equations of motion involve multipolar orders through $\vc{N}$ and $\bsym{\Psi}$, which are defined by (\ref{3Nspin}) and (\ref{3Psi}). In contrast with the dipole order presented in the Corinaldesi-Papapetrou context, the equations (\ref{TDspin}) and (\ref{TDmoment}) show spin-momentum couplings. However, these kind of contributions remain implicit in (\ref{CPspin}) and (\ref{CPmoment1}) through (\ref{3masa2}) with (\ref{CPmoment2}).\\

As a consequence of the equations presented in this paper, an quantitative analisys of the multipolar structure effects could be made. However, this initiative would require to state a model for the mass-energy distribution of the body, which is out of the scope of this paper. Related studies have been performed under particular quadrupolar models by Bini \cite{bini} and Steinhoff \cite{steinhoff2}.

\section*{Appendix}

In the Dixon's scheme the multipole moments of the energy-momentum tensor of an extended body are denoted by $I^{\gamma_1\dots\gamma_n\mu\nu}$ with the following symmetries relations

\begin{eqnarray}\label{GRmoment1}
  I^{\gamma_1\dots\gamma_n\mu\nu} &=&I^{(\gamma_1\dots\gamma_n)(\mu\nu)},\quad \textrm{for\ } n\geq0 \nonumber\\
  I^{(\gamma\mu\nu)} &=&0 \qquad \textrm{and}\qquad  I^{(\gamma_1\dots\gamma_n\mu)\nu}=0,\quad \textrm{for\ } n\geq2.
\end{eqnarray}

In his work, Dixon also writes the equations of motion in terms of a different set of moments, termed $J$ by him and defined by

\begin{equation}\label{GRmoment4}
 I^{\kappa_1\dots\kappa_n\lambda\mu}=\frac{4(n-1)}{n+1}J^{(\kappa_1\dots\kappa_{n-1}|\lambda|\kappa_n)\mu},\quad \textrm{for\ } n\geq2.
\end{equation}

These moments represent a relativistic equivalent of the Newtonian multipole moments. They are related to the spacetime structure, through the Synge's world function and the Jacobi propagators, and the mass-energy distribution of the body (see \cite{dixon5}).

In the equations of motion the moments lead to definitions for the force and the torque, given by

\begin{equation}\label{GRforce}
F^\nu = \frac{1}{2}g^{\mu\nu}\sum_{n=2}^N\frac{1}{n!}I^{\gamma_1\dots\gamma_n\alpha\beta}(s)\nabla_\mu g_{\alpha\beta,\gamma_1\dots\gamma_n}(z)
\end{equation}

and

\begin{equation}\label{GRtorque}
 L^{\kappa\lambda} = \sum_{n=1}^{N-1}\frac{1}{n!}g^{\eta[\kappa}I^{\lambda]\gamma_1\dots\gamma_n\alpha\beta}g_{\{\eta\beta,\alpha\}\gamma_1\dots\gamma_n}.
\end{equation}

In (\ref{GRtorque})

\begin{equation}
g_{\{\eta\beta,\alpha\}}\equiv g_{\eta\beta,\alpha}-g_{\beta\alpha,\eta}+g_{\alpha\eta,\beta}.
\end{equation}

It can be seen from (\ref{GRforce}) and (\ref{GRtorque}) that the force and the torque arise from a coupling between the structure of the body and the spacetime, the latter represented in the metric. When the body loses its test characteristic, $F^\nu$ and $L^{\kappa\lambda}$ get the so-called self-force and self-torque contributions, by virtue of the influence of the mass-energy distribution into the spacetime structure. 

\end{document}